\documentclass[showpacs,preprintnumbers,amsmath,amssymb]{revtex4}
\usepackage{setspace}
\usepackage{graphicx}
\begin{document}

\title{Collective Properties of Excitons in Presence of a Two-Dimensional Electron Gas}

\author{ Oleg L. Berman$^1$, Godfrey Gumbs$^2$, and Patrick A. Folkes$^3$ }
\affiliation{\mbox{$^1$Physics Department, \\ New York City College of Technology of the
City University of New York} \\ 300 Jay Street, Brooklyn, NY 11201 \\
\mbox{$^2$ Department of Physics and Astronomy, \\
Hunter College of the City University of New York} \\
695 Park Avenue, New York, NY 10065 \\
\mbox{$^3$ Army Research Laboratory,   2800 Powder Mill
Road,  Adelphi, Maryland 20783-1197}}

\author{}
\affiliation{}

\date{\today}

\begin{abstract}
We have studied the collective properties of two-dimensional (2D)
excitons  immersed  within a quantum well which contains 2D excitons
and a two-dimensional electron gas (2DEG). We have also
analyzed the excitations for a system of 2D dipole excitons with
spatially separated electrons and holes in a pair of
quantum wells (CQWs)  when one of the wells contains a 2DEG.
Calculations of the superfluid density and the Kosterlitz-Thouless
(K-T) phase transition temperature for the 2DEG-exciton
system in a quantum well have shown that the K-T
transition temperature increases with increasing exciton
density and that it might be possible  to have fast long
range transport of excitons. The superfluid density and
the K-T transition temperature for dipole excitons in CQWs
in the presence of a 2DEG in one of the wells increases
with increasing inter-well separation.

\vspace{0.1cm}



\end{abstract}

\maketitle

 The great interest to the dipole excitons in coupled quantum wells (CQWs) with  spatially separated electron QW and hole QW interest  has been stimulated  by the possibility
of Bose-Einstein condensation (BEC) and the superfluidity of
dipole excitons formed from electron-hole (e-h)  pairs. These may
result in persistent electrical currents in each QW or coherent
optical properties and Josephson junction phenomena~\cite{Lozovik,Shevchenko,Berman,Poushnov,Berman_Tsvetus,Ovchinnikov,BLSC}.
The great experimental success was achieved now in this field~\cite{Snoke,Butov,Eisenstein,Timofeev}.

The coupled QW  system is conceptually simple: negative
electrons are trapped in a two-dimensional plane, while an equal
number of positive holes is trapped in a parallel plane a distance
$D$ away. In this system, the electron
and hole wavefunctions have very little overlap, so that the
excitons can have very long lifetime  ($> 100$ ns), and therefore
they can be treated as metastable particles to which
quasiequlibrium statistics apply. Also, when $D$ is large enough,
the interactions between the excitons are entirely dipole-dipole
repulsive.

In this paper, we consider the collective properties of (a) excitons within one
QW in the presence of a 2D electron gas (2DEG) and (b) dipole excitons
with spatially separated electrons and holes in CQWs when the number of
electrons in the electron well is much larger than the number of holes in
the hole well. In case (b),  the system can be treated as a system of dipole
excitons in the presence of a 2DEG in one well. The
photoluminescence (PL) spectra caused by  the recombination of 2D
electrons  in a single heterojunction quantum well (SHQW)  result in the experimental observation of the excitons coexisting
with a degenerate 2DEG in the same subband \cite{Folkes1}. As shown by time-resolved PL
measurements,  an increase in the integrated PL intensity and the anomalously fast occurrence of PL from the
region around the filament is induced by the screening response of the 2DEG-exciton
system to the abrupt appearance of a remote photocurrent filament in
the 2DEG \cite{Folkes3}. These observations in
conjunction with theory reported in this paper suggest the
occurrence of fast long-range transport of excitons.

We emphasize, that although the K-T transition has been investigated for 2D excitons \cite{BLSC,Snoke},
the key result of our Paper is that we include the 2DEG in our considerations. Consequently, the resulting formulas for
the superfluid density K-T transition temperature are a novel result which we present below. We then try to connect our theory
 with recent experimental results as an alternative explanation of the data.

We now describe the theoretical calculations for determining the
superfluid density and  K-T transition temperature in a QW
containing  2D excitons along with the 2DEG. There are two types of
interaction in this system: exciton-exciton ($X-X$) and
exciton-electron ($X-e$) interaction, while in the absence of 2DEG
there is only $X-X$ interaction. In a QW in the presence of 2DEG the
$X-e$ interaction plays an important role, since $X-e$ scattering is
twice as efficient as $X-X$ scattering~\cite{Koch}. The spectrum of
collective excitations (quasiparticles) in a QW containing the 2D
excitons has the sound branch. This sound branch on the spectrum of
collective excitations is indicative of the interaction strength of
exciton-exciton ($X-X$) scattering and the energy spectrum
$\varepsilon_{X-X}(p)$  in the framework of weakly non-ideal dilute
Bose gas provided by Bogoliubov approximation
\cite{Abrikosov,Griffin} is given by $\varepsilon_{X-X}(p) =
c_{s}p$, where $c_{s} = U_{0}n/M$, $p$ is momentum, $c_{s}$ is the
sound velocity of collective branch due to $X-X$ scattering, $n$ is
the exciton density, $M = m_{e} + m_{h}$ is the exciton mass
($m_{e}$ and $m_{h}$ are the effective masses of electron and hole,
respectively), $U_{0}$ is the zero-order Fourier-component (at
$p=0$) of the exciton-exciton repulsion potential assuming a hard
sphere approximation, given by \cite{Berman_Lozovik_Snoke} $U_{0} =
6 e^{2} a_{2D}/\epsilon$, where $e$ is the electron charge, $a_{2D}
= \hbar^{2}\epsilon/(2\mu_{e-h}e^{2})$, $\mu_{e-h} =
m_{e}m_{h}/(m_{e} + m_{h})$, $\epsilon$ is the dielectric constant.

For the system of the spatially separated excitons and electrons in
coupled quantum wells (CQWs) it was shown that the effective $X-X$
interaction pair potential becomes weaker due to screening by
electrons~\cite{Kulakovskii}. We take into account this screening
effect by introducing the following phenomenological form of the
 zero-order Fourier-component  of the exciton-exciton repulsion potential
in the presence of the 2DEG: $\tilde{U}_{0} = \beta(n,n_{e}) U_{0}$,
where $\beta(n,n_{e}) < 1$ is the coefficient reflecting the
electron screening effects, and $n_{e}$ is the electron density.
This substitution causes the electron screening renormalization of
the sound velocity $\tilde{c}_{s} = \beta(n,n_{e}) c_{s}$. Note that
$\beta(n,n_{e})$ can depend on the exciton density $n$ and electron
density $n_{e}$. The calculation of a more accurate phenomenological
relationship will be addressed in future work.

The  superfluid-normal phase transition in the 2D exciton system is
the K-T transition \cite{Kosterlitz}. The transition
temperature $T_{\rm c} $, for the K-T transition to
the superfluid state  in a 2D exciton system is determined by the
equation \cite{Kosterlitz}\  $T_{\rm c} = \pi \hbar^2 n_{\rm s}
(T_{\rm c})/{2 k_B M}$, where $n_{\rm s} (T)$ is the superfluid
density of the exciton system as a function of temperature $T$  and
$k_B$ is Boltzmann' constant.

The function $n_{\rm s} (T)$ has been calculated from the
relation $n_{\rm s} = n - n_n$  ($n$ is the total exciton
density, $n_{n}$ is the normal component
density). We determine the normal component density by means of the usual
procedure \cite{Abrikosov,Berman_Lozovik_Gumbs}. Let us suppose that the
exciton system  moves with  velocity $\mathbf{u}$. At finite
temperature $T$, dissipating quasiparticles will appear in this
system. Since their density is small at low temperatures, one can
assume that the gas of quasiparticles is an ideal Bose gas. To
calculate the superfluid component density, we find the total current
of quasiparticles in a
 frame in which the superfluid component is at rest.
Then we obtain the mean total current of 2D excitons in the
coordinate system, moving with a velocity ${\bf u}$:

\begin{eqnarray}
\label{nnor} && \left\langle \mathbf{J} \right\rangle = \frac{1}{M}
\left\langle \mathbf{p} \right\rangle =\frac{1}{M} s \int_{}^{}
\frac{d\mathbf{p}}{(2\pi \hbar)^{2}} \mathbf{p} \nonumber \\
&& \times f\left[\varepsilon_{X-X} (p) - \mathbf{p}\mathbf{u}
\right]
 \ ,
\end{eqnarray}
where $f\left[\varepsilon_{X-X} (p)\right] =
\left(\exp\left[\varepsilon_{X-X} (p)/(k_{B}T)\right] -
1\right)^{-1}$ is the Bose-Einstein distribution function, $s$ is
the level degeneracy (equal to $4$ for excitons in GaAs quantum
wells). Expanding the expression inside the integral in the first
order by $\mathbf{p}\mathbf{u}/(k_{B}T)$, we have:

\begin{eqnarray}\label{J_Tot}
 && \langle \mathbf{J} \rangle = -
\frac{\mathbf{u}}{2M} s \int\frac{d\mathbf{p}}{(2\pi
\hbar)^{2}}p^{2}\frac{\partial f\left[\varepsilon_{X-X}
(p)\right]}{\partial \varepsilon_{X-X}}  \nonumber \\
&& =   \frac{3 \zeta (3) }{2 \pi
\hbar^{2}}\frac{k_{B}^{3}T^3}{M}\frac{1}{\tilde{c}_{s}^{4}}
\mathbf{u} \ ,
\end{eqnarray}
where $\zeta (z)$ is the Riemann zeta function ($\zeta (3) \simeq
1.202$). Then we define the normal component density $n_{n}$
as\cite{Abrikosov} $\langle  \mathbf{J} \rangle = n_n  \mathbf{u}$.
Comparing this definition of $n_{n}$ and\ (\ref{J_Tot}),  we obtain
the expression for the normal density $n_{n}$. Consequently, we have
for the superfluid density:

\begin{eqnarray}
\label{n_s} n_{\rm s} = n - n_n = n -
 \frac{3 s \zeta (3) }{2 \pi \hbar^{2}}
\frac{k_{B}^{3}T^3}{\tilde{c}_{s}^{4}} \ .
\end{eqnarray}

In a 2D system, superfluidity of excitons in a 2DEG
appears below the K-T  transition temperature, where
only coupled vortices are present \cite{Kosterlitz}.
Making use of Eq.\ (\ref{n_s}) for the density
$n_{s}$ of the superfluid component, we obtain an
equation for the Kosterlitz-Thouless transition
temperature $T_{\rm c}$. Its solution is given by

\begin{eqnarray}
\label{tct} && T_{\rm c} = \left[\left( 1 +
\sqrt{\frac{32}{27}\left(\frac{M k_{B}T_{\rm c}^{0}}{\pi \hbar^{2}
n}\right)^{3} + 1} \right)^{1/3}   \right.  \nonumber \\ && - \left.
\left( \sqrt{\frac{32}{27} \left(\frac{M k_{B}T_{\rm c}^{0}}{\pi
\hbar^{2} n}\right)^{3} + 1} - 1 \right)^{1/3}\right]
\frac{T_{\rm c}^{0}}{ 2^{1/3}} \  .
\end{eqnarray}
Here $T_{\rm c}^{0}$ is an auxiliary quantity, equal to the temperature
at which the superfluid density vanishes in the mean-field
approximation (i.e., $n_{s}(T_{\rm c}^{0}) = 0$),

\begin{equation}
\label{tct0} T_{\rm c}^0 = \frac{1}{k_{B}} \left( \frac{2 \pi
\hbar^{2} n c_{s}^{4} M}{3 s \zeta (3)} \right)^{1/3} \ .
\end{equation}
The temperature $T_{\rm c}^{0}$ may be used as a crude estimate of the
crossover region where local superfluid density appears for rare
exciton system in 2DEG on a scale smaller or of the order of the mean
intervortex separation in the system. The local superfluid density
can manifest itself in local optical properties or local transport
properties.

The K-T  transition temperature $T_{\rm c}$ obtained in
our calculations as a function of the exciton density $n$ in a single quantum well is
given by Eq.~(\ref{tct}). It is shown that $T_{\rm c}$ increases when
the exciton density $n$ increases. Besides,  at the fixed temperature the superfluidity in a single quantum well
 exists at the exciton densities greater than the critical one. This critical exciton density increases when the temperature increases.
The theoretical results also show that a superfluid transition at
$T_{\rm c} = 2 \ \mathrm{K}$ requires an exciton density
of $7.5 \times 10^{10} \ \mathrm{cm}^{-2}$; which is reasonably
 close to the theoretical estimate of the exciton density realized in the experiment of Ref.~[\onlinecite{Folkes3}].

Besides, we investigate   dipole excitons with
 spatially separated electrons and holes in CQWs when
the  number of electrons in the electron well is much more than the
number of  holes in the hole well. In a dilute system when the
numbers of electrons and holes are equal, ($na_{2D}^{2} \ll 1$) two
dipole excitons repel each other like  two parallel dipoles with the
pair repulsion potential $U(R) = e^{2}D^{2}/(\epsilon R^{3})$, where
$R$ is the distance between two excitons, and $D$ is the inter-well
septation. The collective excitations in this  spectrum have the
sound spectrum \cite{Yudson}: $\varepsilon_{X-X}(p) = c_{s}(D)p$.
The sound velocity $c_{s}(D)$, when the electron and hole numbers
are equal, is calculated in the ladder approximation assuming the
vertex correction is  equal the sum of the ladder diagrams and given
by \cite{BLSC,Yudson}

\begin{equation}\label{c1D}
c_{s}(D) =  \left(\frac{4\pi \hbar^{2}n}{M^{2} \log \left(
\frac{\epsilon^{2}\hbar^{4}}{8\pi s^2 n M^2 e^4 D^4} \right)}
\right)^{1/2} \ .
\end{equation}

For the case when number of electrons is much greater than number of
holes, taking into account the electron screening effects we assume
the following phenomenological form of the exciton-exciton dipole
repulsion potential in the presence of the 2DEG: $\tilde{U} (R) =
 \gamma(n,n_{e},D) e^{2}D^{2}/(\epsilon R^{3})$, where
$\gamma (n,n_{e},D) < 1$ is the coefficient reflecting the electron
screening effects. This substitution results in the electron
screening renormalization of the sound velocity:

\begin{equation}\label{c1Dr}
\tilde{c}_{s}(D) =  \left(\frac{4\pi \hbar^{2}n}{M^{2} \log \left(
\frac{\epsilon^{2}\hbar^{4}}{8\pi s^2 n M^2 \gamma^{2}(n,D) e^4 D^4}
\right)} \right)^{1/2} \ .
\end{equation}

It follows from Eq.~(\ref{c1Dr}) that $\tilde{c}_{s}(D) < c_{s}(D)$
due to the fact that $\gamma (n,n_{e},D) < 1$. Therefore, the sound
velocity for the collective excitation spectrum in the CQWs with the
number of electrons greater than number of holes can be represented
as  $\tilde{c}_{s}(D) < \theta (n,n_{e},D)c_{s}(D)$, where $\theta
(n,n_{e},D)<1$ is the phenomenological parameter. The rest of the
procedure for the calculation of the superfluid density and
Kosterlitz-Thouless temperature is the same as for the single
quantum well in the presence of 2DEG. The superfluid density for
CQWs can be calculated by applying Eq.~(\ref{n_s}) and and
Kosterlitz-Thouless temperature can be calculated by using
Eq.~(\ref{tct}) with the substitution of the renormalized sound
velocity from Eq.~(\ref{c1Dr}) instead of $c_{s} = \beta
(n,n_{e})U_{0}n/M$ for the single quantum well.

The calculation of the phenomenological parameters $\beta(n,n_{e})$
for a single quantum well in the presence of 2DEG and
 and $\theta (n,n_{e},D)$ for CQWs with the  number of electrons higher than the number of holes is
 an open question which should be answered by future work.

\begin{figure}[t] 
   \centering
  \includegraphics[width=6.0in]{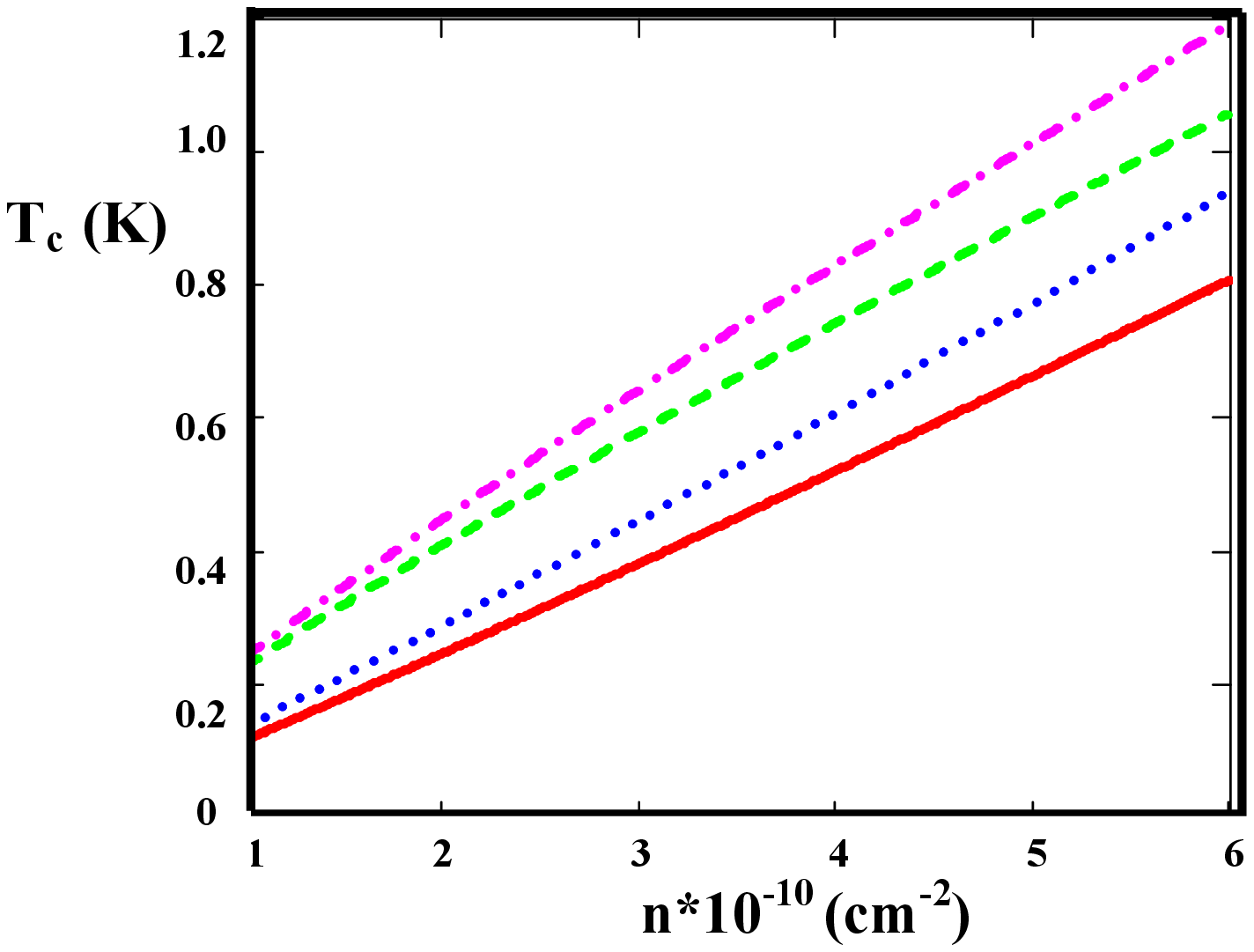}
   \caption{(Color on line)  Kosterlitz-Thouless transition temperature $T_{\rm c}$
as a function of the exciton density $n$  in CQWs for different
inter-well separations $D$ and phenomenological parameter $\theta
(n.n_{e},D)$: $D= 0.5 \ \mathrm{nm}$, $\theta (n.n_{e},D) = 0.5$ --
solid curve; $D= 0.5 \ \mathrm{nm}$,  $\theta (n.n_{e},D) = 0.94$ --
dotted curve; $D= 5 \ \mathrm{nm}$,  $\theta (n.n_{e},D) = 0.5$
 -- dashed curve; $D= 5 \ \mathrm{nm}$, $\theta (n.n_{e},D) = 0.94$  --
dashed-dotted curve.  }
   \label{TCD}
\end{figure}

The K-T  transition temperature $T_{\rm c}$ obtained in
our calculations as a function of the exciton density $n$ in CQWs is
presented in Fig.\ \ref{TCD}. It is shown that $T_{\rm c}$
increases monotonically with increasing exciton density $n$
 and the inter-well separation $D$, which can be explained by
the fact that the dipole-dipole repulsion causing the superfluidity
increases if $D$ increases. Furthermore, the results of our
calculations show that $T_{\rm c}$ increases when  $\beta(n,n_{e})$
for a single well or  $\theta (n,n_{e},D)$ for CQWs increase.
According to Fig.\ \ref{TCD},  analogously to the exciton-electron
system a single  quantum well at the fixed temperature the
superfluidity in CQWs exists at the exciton densities greater than
the critical one.
  This critical exciton density increases, when the temperature increases and the interwell separation $D$ decreases.
  Increasing $\beta(n,n_{e})$ for a single quantum well or $\theta (n,n_{e},D)$ for CQWs
    results in the decreasing  critical exciton density.

 An experimental observation  \cite{Folkes3} has been reported on a sharp
increase in the average photocurrent at the critical voltage $V_{\rm g}
= - 36 \ \mathrm{V}$ and hysteresis in the I-V characteristics with
subsequent variation of $V_{\rm g}$  which is indicative of the
formation of a photocurrent filament \cite{Scholl} from the undoped
GaAs layer into the 2DEG in the SHQW. It has been observed in  Ref.~[\onlinecite{Folkes3}] that there is  concurrent, significant changes in the time-integrated excitonic PL at  $V_{\rm g} =
- 36\  \ \mathrm{V}$. The data show that shortly after excitation at $V_{\rm g} = -36 \
\mathrm{V}$, the PL intensity at the filament position is
approximately a factor of 8 larger than the PL intensity observed at
the same point and time with $V_{\rm g} = 0 \ \mathrm{V}$ even though
the residual laser excitation intensity at the filament position is
unchanged. TR PL profile data \cite{Folkes3} shows that there is no
observed delay between the PL peaks in the photoexcitation and the
filament regions indicating that no in-plane transport of excitons from
the excitation region to the filament region is observed. However,
the transport of excitons in a time less than $0.1 \ \mathrm{ns}$ (the
experimental resolution time) cannot be ruled out. The data suggests
that an exciton excitation energy transport mechanism is responsible
for the fast observation of $1550 \ \mathrm{meV}$ PL from the filament region.
One possibility is that in response to the filament-induced
potential, excitons undergo a Kosterlitz-Thouless superfluid
transition \cite{Kosterlitz} with a long-range transport of exciton
excitation energy. Theory points out that exciton superfluidity does
not involve transport of mass; instead it involves a transport of
excitation energy \cite{Hanamura}. The filament-induced increase in
exciton density and lifetime along with efficient electron-exciton
scattering \cite{Malpuech} in the presence of the 2DEG are conducive
to exciton condensation. Using
theoretical results \cite{Mahan2} we estimate   that the
peak exciton density in the SHQW is approximately $4\times 10^{10} \
\mathrm{cm}^{-2}$. There is a large uncertainty in calculating and
measuring the actual exciton density in a QW with a 2DEG-exciton
system. Nevertheless, the observed appearance of high-intensity PL
at a point $25 \ \mathrm{\mu m}$ from the edge of the excitation
region less than $0.1 \ \mathrm{ns}$ after excitation is a possible
manifestation of a long range transport of exciton excitation
energy. Further research is needed to understand this exciton
excitation energy transport mechanism and its potential application
to optoelectronic devices.

The experimental results obtained from measurement of the exciton
dephasing rate show that electron-exciton scattering is twice as
efficient as exciton-exciton scattering. It is important to note
that the enhanced electron-exciton scattering also leads to
efficient relaxation of exciton energy in agreement with previous
theoretical results~\cite{Malpuech}.  Efficient exciton energy
relaxation involves collective plasmon-phonon interactions in the
exciton/two-dimensional electron gas (exciton/2DEG) system which
could result in the formation of the second sound branch.
Experimental results~\cite{Folkes3}  clearly show that the
collective screening response of the exciton/2DEG system to a
photo-induced potential results in fundamental changes in exciton
characteristics including the possibility of condensation and
transport of excitons. The 2DEG density in our experiment is
$2\times 10^{11} \ \mathrm{cm}^{-2}$.

  Negatively charged excitons (trions), which consist of two
electrons bound to a single hole with a theoretical~\cite{Stebe} and
observed~\cite{Finkelstein,Buhmann,Shields}  binding energy relative
to the exciton of approximately $1 \ \mathrm{meV}$ have been
observed in modulation-doped GaAs QWs. A photoluminescence peak from
trions has been observed from the exciton \ 2DEG system in our
structure around $1 \ \mathrm{meV}$ below the exciton PL peak only
when the exciton linewidth has been minimized using slight
variations of the applied gate voltage~\cite{Folkes_4}.   In the
present experiments, broadening of the PL linewidths at $V_{g} = -36
\ \mathrm{V}$ and the short exciton lifetime of $0.5 \ \mathrm{ns}$
at $V_{g} = 0$ precluded the resolution of a PL peak from trions.

In conclusion, calculations of the superfluid density and the K-T  transition
temperature for the 2DEG-exciton system show that the K-T transition temperature
increases with increasing exciton density and support the possible occurrence of a
fast transport of exciton excitation energy. The superfluid density and the K-T transition
temperature for dipole excitons in CQWs in the presence of a 2DEG in one of the
 wells increases with increasing inter-well separation $D$. The superfluid density and
the K-T transition temperature  in CQWs increase when the
phenomenological screening parameters $\beta(n,n_{e})$ for a single
well or  $\theta (n,n_{e},D)$ for CQWs in the presence of 2DEG
increase.

\vskip 0.1in

\noindent Acknowledgments:  O.~L.~B. would like to acknowledge the
support by PSC CUNY grant 621360040,  GG would like to acknowledge
the support by contract  FA9453-07-C-0207 of AFRL.

\newpage

\end{document}